\documentclass[conference]{IEEEtran}
\usepackage[T1]{fontenc}
\usepackage[utf8]{inputenc}
\usepackage{pgfplots}
\usepackage[binary-units=true]{siunitx}
\DeclareSIUnit{\bits}{bits}
\usepackage{amssymb}
\usepackage{subcaption} 
\usepackage{amsmath}
\usepackage{amsthm}
\usepackage{mathtools}
\usepackage{array}
\usepackage[noadjust]{cite}
\usepackage{tikz}
\usepackage[bookmarks=false]{hyperref}
\usepackage[acronym]{glossaries}
\usepackage{hhline}
\usepackage{multirow}
\usepackage{siunitx}
\DeclareUnicodeCharacter{2212}{−}
\usepgfplotslibrary{groupplots,dateplot}
\usetikzlibrary{patterns,shapes.arrows}
\pgfplotsset{compat=newest}
\usepackage[ruled,vlined]{algorithm2e}
\usepackage{todonotes}
\newcommand{\comment}[1]{}

\DeclareSIUnit{\kmph}{kmph}
\newacronym{3GPP}{3GPP}{3rd Generation Partnership Project}
\newacronym{5GNR}{5G NR}{5G New Radio}

\newacronym{AWGN}{AWGN}{additive white Gaussian noise}

\newacronym{BCE}{BCE}{binary cross-entropy}
\newacronym{BER}{BER}{bit error rate}
\newacronym{BICM}{BICM}{bit-interleaved coded modulation}
\newacronym{BLER}{BLER}{block error rate}
\newacronym{BMDR}{BMDR}{bit-metric decoding rate}

\newacronym{CDF}{CDF}{cumulative distribution function}
\newacronym{CER}{CER}{codeword error rate}
\newacronym{CNN}{CNN}{convolutional neural network}
\newacronym{Conv2D}{Conv2D}{convolutional}
\newacronym{CSI}{CSI}{channel state information}
\newacronym{CP}{CP}{cyclic prefix}
\newacronym{CvT}{CvT}{convolutional-vision-transformer}

\newacronym{eMBB}{eMBB}{enhanced mobile broadband}

\newacronym{FC}{FC}{fully-connected}
\newacronym{FFT}{FFT}{fast Fourier transform}

\newacronym{HARQ}{HARQ}{hybrid automatic repeat request}

\newacronym{IFFT}{IFFT}{inverse fast Fourier transform}
\newacronym{iid}{i.i.d.\@}{independent and identically distributed}
\newacronym{ISI}{ISI}{inter-symbol interference}

\newacronym{K}{K}{keys}
\newacronym{KL}{KL}{Kullback–Leibler}

\newacronym{LDPC}{LDPC}{low-density parity-check}
\newacronym{LMMSE}{LMMSE}{linear minimum mean square error}
\newacronym{LLR}{LLR}{log-likelihood ratio}
\newacronym{LOS}{LOS}{line-of-sight}
\newacronym{LS}{LS}{least square}

\newacronym{MAP}{MAP}{maximum a posteriori}
\newacronym{MCS}{MCS}{modulation and coding scheme}
\newacronym{MHA}{MHA}{multi-head-attention}
\newacronym{MI}{MI}{mutual-information}
\newacronym{MIMO}{MIMO}{multiple-input multiple-output}
\newacronym{ML}{ML}{maximum-likelihood}
\newacronym{MU-MIMO}{MU-MIMO}{multi-user MIMO}
\newacronym{MSE}{MSE}{mean squared error}

\newacronym{NN}{NN}{neural network}

\newacronym{OFDM}{OFDM}{orthogonal frequency division multiplexing}

\newacronym{PRB}{PRB}{physical resource block}
\newacronym{PUSCH}{PUSCH}{physical uplink shared channel}

\newacronym{Q}{Q}{queries}
\newacronym{QAM}{QAM}{quadrature amplitude modulation}
\newacronym{QPSK}{QPSK}{quadrature phase-shift keying}

\newacronym{RE}{Rs}{resource elemens}
\newacronym{REs}{REs}{resource elements}
\newacronym{ReLU}{ReLU}{rectified linear unit}

\newacronym{SGD}{SGD}{stochastic gradient descent}
\newacronym{SINR}{SINR}{signal-to-interference noise ratio}
\newacronym{SNR}{SNR}{signal-to-noise ratio}

\newacronym{wrt}{w.r.t.\@}{with respect to}

\newacronym{ZF}{ZF}{zero-forcing}
\newacronym{RL}{RL}{reinforcement-learning}

\newacronym{uRLLC}{uRLLC}{ultra-reliable and low-latency communications}

\newacronym{V}{V}{values}

\renewcommand{\vec}[1]{\mathbf{#1}}

\newcommand{\hv}{\vec{h}}

\newcommand{\nv}{\vec{n}}

\newcommand{\wv}{\vec{w}}
\newcommand{\xv}{\vec{x}}
\newcommand{\yv}{\vec{y}}
\newcommand{\zv}{\vec{z}}

\newcommand{\Am}{\vec{A}}

\newcommand{\Hm}{\vec{H}}
\newcommand{\Id}{\vec{I}}

\newcommand{\Km}{\vec{K}}
\newcommand{\Lm}{\vec{L}}

\newcommand{\Qm}{\vec{Q}}
\newcommand{\Rm}{\vec{R}}

\newcommand{\Vm}{\vec{V}}
\newcommand{\Wm}{\vec{W}}
\newcommand{\Xm}{\vec{X}}

\newcommand{\Zm}{\vec{Z}}

\newcommand{\Gc}{{\cal G}}

\newcommand{\Qc}{{\cal Q}}

\newcommand{\CC}{\mathbb{C}}

\newcommand{\RR}{\mathbb{R}}

\newcommand{\LB}{\left(}
\newcommand{\RB}{\right)}

\newcommand{\LSB}{\left[}
\newcommand{\RSB}{\right]}

\renewcommand{\exp}[1]{\mathop{\mathrm{exp}}\LB #1\RB}

\newcommand{\EE}{{\mathbb{E}}}

\newlength{\dhatheight}

\IEEEoverridecommandlockouts
\begin{document}
\begin{NoHyper}
\title{Convolutional Self-Attention-Based Multi-User MIMO Demapper}

\author{
\IEEEauthorblockN{Athur Michon\IEEEauthorrefmark{1}, Fayçal Ait Aoudia\IEEEauthorrefmark{2}, K. Pavan Srinath\IEEEauthorrefmark{3}\thanks{This work was carried out when A. Michon and F. Ait Aoudia were at Nokia Bell Labs France.}}
\IEEEauthorblockA{\IEEEauthorrefmark{1}INP-ENSEEIHT, 31000 Toulouse, France}
\IEEEauthorblockA{\IEEEauthorrefmark{2}NVIDIA, 06906 Sophia Antipolis, France}
\IEEEauthorblockA{\IEEEauthorrefmark{3}Nokia Bell Labs France, 91620 Nozay, France \\
arthur.michon@hotmail.fr, faitaoudia@nvidia.com, pavan.koteshwar\_srinath@nokia.com
}}

\maketitle

\begin{abstract}

In \gls{OFDM}-based wireless communication systems, the \gls{BER} performance is heavily dependent on the accuracy of channel estimation. It is important for a good channel estimator to be capable of handling the changes in the wireless channel conditions that occur due to the mobility of the users. In recent years, the focus has been on developing complex \gls{NN}-based channel estimators that enable an error performance close to that of a genie-aided channel estimator. This work considers the other alternative which is to have a simple channel estimator but a more complex \gls{NN}-based demapper for the generation of soft information for each transmitted bit. In particular, the problem of reversing the adverse effects of an imperfect channel estimator is addressed, and a convolutional self-attention-based neural demapper that significantly outperforms the baseline is proposed.
\end{abstract}

\glsresetall

\section{Introduction} 
\label{sec:introduction}

With an ever-growing demand for data rate in wireless communications, machine-learning-based alternatives to the various classical functional blocks in the physical layer have been under consideration in recent times. The recent advances in machine learning in several fields (image recognition, natural language processing, etc.) have aroused an interest for the same in the field of wireless communications. It is a common practice nowadays to replace the functional blocks of a transceiver chain like channel estimation, equalization, \gls{LLR} generation, etc., by deep neural networks (DNNs) \cite{8868708,shental2020machine,8272484}. More recent techniques propose to jointly replace a group of functional blocks by DNNs. An example is \cite{honkala2021deeprx}, which proposes to jointly learn channel estimation, equalization, and \gls{LLR} generation (demapping) using a DNN. There have also been several proposals to perform end-to-end learning -- the joint optimization of the transceiver using an auto-encoder \cite{Dorner_2018}.

In many instances, it might not be feasible to have a very complex \gls{NN} in a product due to the large number of trainable parameters. For example, the \gls{NN} of \cite{honkala2021deeprx} has around $1.2$ million parameters for uplink MIMO reception and detection, and this scales with the number of users. So, it might be more practical in some cases to use a simple channel estimator like the \gls{LMMSE} channel estimator at pilot locations, and employ a more robust demapper to alleviate the adverse impact of the imperfect channel estimator. A very simple \gls{MU-MIMO} receiver would perform channel estimation on the \gls{REs} at the pilot locations, use the nearest-pilot-location estimate for other data-carrying \gls{REs}, perform equalization using the \gls{LMMSE} detector, and then use a Gaussian demapper to generate the \gls{LLR} for each transmitted bit of each user. However, this would entail a significant drop in the \gls{BER} performance. 

In this paper, we focus on replacing only the Gaussian demapper with a learned \gls{NN} demapper with the goal of reversing the effects of imperfect channel estimation. The idea is to capture the time-frequency correlation between the equalization errors in an \gls{OFDM} grid, and for this purpose, we make use of a \gls{CNN}. Since there is an interdependency between the equalization errors of all users due to inter-user interference, we propose to deal with it by making use of self-attention mechanism \cite{vaswani2017attention}. The most significant difference between this work and \cite{goutay} is that the latter considers a more complex LMMSE channel estimator which requires the spatial, temporal, and frequency correlation matrices (see \cite[Sec. II-B]{goutay}) of the channel for each user, and this has a lot of practical limitations.  

\begin{figure*}[t!]
    \centering
    \includegraphics[width=1.\textwidth]{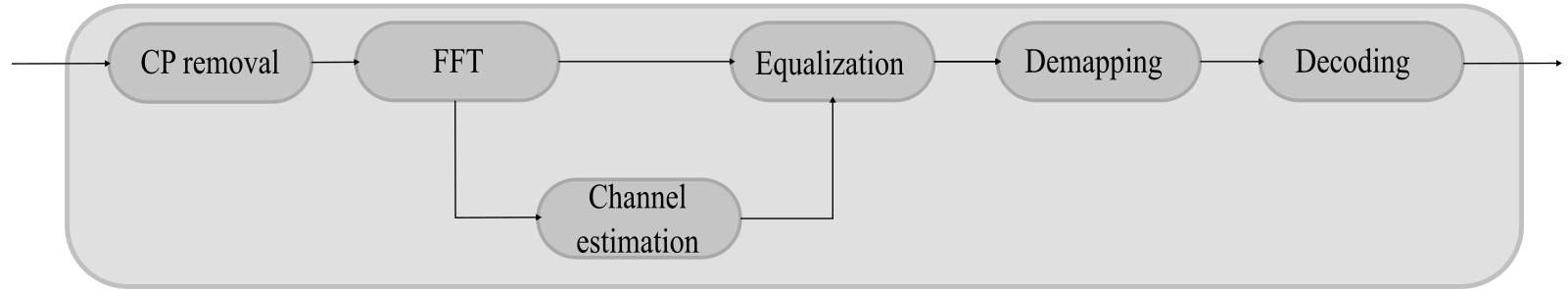}
    \caption{Block diagram of the receiver.}
    \label{fig:receiver}
\end{figure*}

The rest of the paper is organized as follows. Section \ref{sec:problem_positioning} describes the system model and the set-up while Section \ref{sec:NN} presents the proposed neural demapper. Simulation results are provided in Section \ref{sec:evaluations} and the concluding remarks constitute Section \ref{sec:conclusion}.

{\it Notation}: The field of complex numbers and real numbers are respectively denoted by $\CC$ and $\RR$. Throughout the rest of the paper, boldface uppercase (lowercase) letters denote matrices (vectors). The Frobenius norm of a matrix is denoted by $\Vert \Xm \Vert $. The complex conjugate of $x$ is denoted by $x^*$, and the transpose and Hermitian transpose of a matrix $\Xm$ by $\Xm^T$ and $\Xm^H$, respectively. The identity matrix is denoted by $\Id$ with its size understood from the context.

\section{Preliminaries} 
\label{sec:problem_positioning}

\subsection{System Model}
\label{sec:system_model}

In this paper, we consider multiple single-antenna users transmitting on the uplink, but a similar treatment holds for \gls{MU-MIMO} transmission on the uplink/downlink as well. The transceiver model is based on the \gls{3GPP} recommendations for 5G \gls{PUSCH}, and the receiver architecture is illustrated in Fig. \ref{fig:receiver}. 

We assume that there are $N_u$ single-antenna users transmitting to a base station equipped with $N_r$ receive antennas. At the transmitter of each user, message bits are encoded using \gls{LDPC} coding and then independently mapped to constellation symbols from a unit-energy constellation denoted by $\Qc$ (for simplicity, we assume the same constellation for all users, but these can be different in practice). These data symbols are then mapped to the data-carrying \gls{REs} and pilot symbols are mapped to pilot-carrying \gls{REs} in the \gls{OFDM} grid \cite[Sec. 6.4]{3GPP_DMRS_2021} before being transmitted over the channel. In this paper, we assume that an \gls{OFDM} grid refers to the grid of \gls{REs} associated with a transmission slot as defined by 3GPP \cite{3GPP_DMRS_2021}, and has $N_f$ subcarriers and $N_t$ symbols (typical value being 14 in 5G NR). The frequency domain signal model for any index pair $(m,n)$ that denotes the $m^{th}$ subcarrier and the $n^{th}$ symbol in the grid is given as
\begin{equation}
    \yv_{m,n} = \Hm_{m,n} \xv_{m,n} + \nv_{m,n}
\end{equation}
where $\yv_{m,n} \in \CC^{N_r \times 1}$ is the received signal vector,  $\Hm_{m,n} \triangleq \LSB \hv_{m,n}^{(1)},\cdots,\hv_{m,n}^{(N_u)} \RSB\in \CC^{N_r \times N_u}$ is the composite channel matrix with $\hv_{m,n}^{(i)} \in \CC^{N_r \times 1}$ representing the channel from User $i$ to the base station, $\xv_{m,n} \in \Qc^{N_u \times 1}$ is the transmitted composite signal from the $N_u$ users, and $\nv_{m,n} \in \CC^{N_r \times 1}$ represents the zero mean \gls{AWGN} with variance $\sigma^2$. Following the definition in \cite[Sec. II-A]{snr_defn}, the instantaneous \gls{SNR} at the receiver for the resource grid in context is defined as
\begin{equation}
    \textrm{SNR} = \frac{\sum_{m,n}\Vert \Hm_{m,n}\Vert^2}{N_f N_t N_r N_u \sigma^2}. 
\end{equation}

\begin{figure*}[htbp]
    \centering
    \includegraphics[width=1.\textwidth]{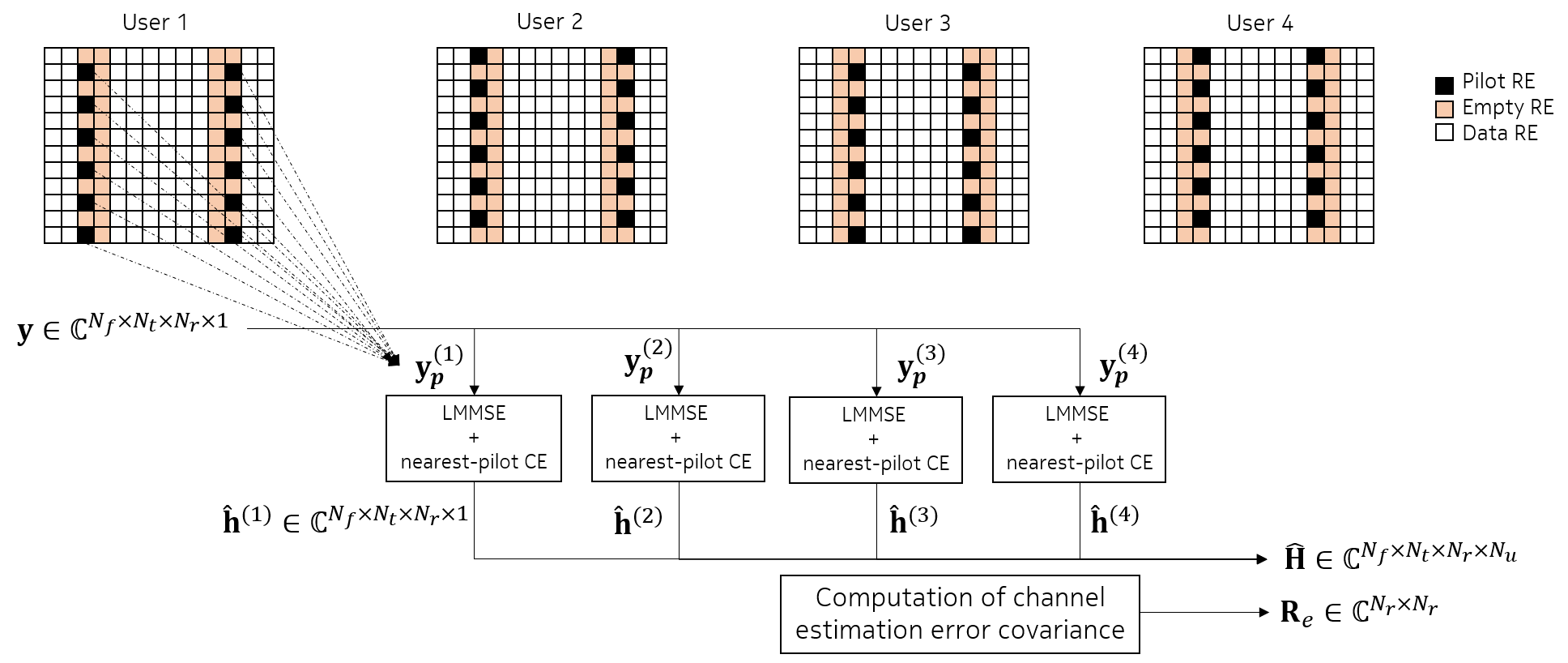}
    \caption{A schematic description of the channel estimation process for $N_u=4$}
    \label{fig:operation}
\end{figure*}

At the base station receiver, after \gls{CP} removal and \gls{FFT} of the received signal, \gls{LMMSE} channel estimation is performed on the signals received at the pilot \gls{REs}. We assume that the pilots of users are transmitted on mutually non-overlapping \gls{REs}, as shown in Fig. \ref{fig:operation}. Therefore, the \gls{LMMSE} channel estimate at any pilot RE with index pair $(m,n)$ for User $i$ is given as
 \begin{equation}
    \hat{\hv}^{(i)}_{m,n} = \LB x^{(i)}_{p}\RB^*\Rm^{(i)}_s\LSB\Rm^{(i)}_s+\sigma^2 \Id\RSB^{-1}\yv_{m,n}, \label{eq:h_hat_1usr} 
\end{equation}
where $x^{(i)}_p$ is the unit-energy pilot signal transmitted by the $i^{th}$ user, and we assume that the spatial correlation covariance $\Rm^{(i)}_S \triangleq \EE\LSB \hv_{m,n}^{(i)}\LB \hv_{m,n}^{(i)} \RB^H \RSB $ (expectation is over the RE indices) and the noise variance are known at the receiver. For the remaining data-carrying \gls{REs}, nearest-pilot (NP) channel estimation is performed, i.e., the channel is estimated to be the same as that of the nearest pilot RE in the resource grid. The composite channel estimate for all users is denoted by $\hat{\Hm}_{m,n}\in \mathbb{C}^{N_r \times N_u}$. The covariance of the channel estimation error at the pilot locations for the $i^{th}$ user is given by the well-known formula as
 \begin{eqnarray} \nonumber
    \Rm^{(i)}_e & \triangleq & \EE\LSB \LB\hv^{(i)} - \hat{\hv}^{(i)}\RB\LB\hv^{(i)} - \hat{\hv}^{(i)}\RB^H \RSB \\
    & = & \Rm^{(i)}_s - \Rm^{(i)}_s\LSB \Rm^{(i)}_s+\sigma^2\Id \RSB^{-1}\Rm^{(i)}_s.
\end{eqnarray}
Therefore, the error covariance for the composite channel estimate is $\Rm_e \triangleq  \EE\LSB \LB\Hm - \hat{\Hm}\RB\LB\Hm - \hat{\Hm}\RB^H \RSB = \sum_{i=1}^{N_u}\Rm^{(i)}_e$. 

With $\hat{\Hm}$ and $\Rm_e$ thus obtained, we have
\begin{eqnarray} \nonumber
    \yv_{m,n}  & = & \hat{\Hm}_{m,n}\xv_{m,n} + \LB \Hm_{m,n}-\hat{\Hm}_{m,n} \RB \xv_{m,n} + \nv_{m,n} \\
    & = & \hat{\Hm}_{m,n}\xv_{m,n} + \wv_{m,n}
\end{eqnarray}
where $\wv_{m,n} \triangleq \LB \Hm_{m,n}-\hat{\Hm}_{m,n} \RB \xv_{m,n} + \nv_{m,n}$ has covariance $\Rm_w \triangleq \Rm_e + \sigma^2 \Id$ and is uncorrelated with $\hat{\Hm}_{m,n}$ (due to LMMSE estimation). Next, noise-whitening is performed as follows. 
\begin{equation}
    \Tilde{\yv}_{m,n} \triangleq \Rm_w^{-\frac{1}{2}}\yv_{m,n} = \Rm_w^{-\frac{1}{2}}\hat{\Hm}_{m,n} \xv_{m,n}+ \Rm_w^{-\frac{1}{2}}\wv_{m,n}.
\end{equation}
The result of this operation is that the covariance of $\Rm_w^{-\frac{1}{2}}\wv_{m,n}$ is the  identity matrix. With $\Tilde{\Hm}_{m,n} \triangleq \Rm_w^{-\frac{1}{2}}\hat{\Hm}_{m,n}$, we perform \gls{LMMSE} equalization to obtain
\begin{equation} \label{eq:lmmse_equalization}
   \hat{\xv}_{m,n} = \Tilde{\Hm}_{m,n}^H(\Tilde{\Hm}_{m,n}\Tilde{\Hm}_{m,n}^H+\Id)^{-1}\Tilde{\yv}_{m,n}.
\end{equation}
The post-equalization error is given by
\begin{equation}
    \zv_{m,n} \triangleq \xv_{m,n} - \hat{\xv}_{m,n} \label{eq:eqNoiseVar}
\end{equation}
with covariance $ \Rm_{z,m,n} \triangleq \Id - \Tilde{\Hm}_{m,n}^H(\Tilde{\Hm}_{m,n}\Tilde{\Hm}_{m,n}^H+\Id)^{-1} \Tilde{\Hm}_{m,n}$.

\subsection{The Gaussian Demapper}
\label{sec:baseline}
From \eqref{eq:lmmse_equalization}, the signal for each user after LMMSE equalization can be written as
\begin{equation}
    \hat{x}_{m,n,i} = x_{m,n,i} + z_{m,n,i}, \forall i=1,\cdots, N_u, \label{eq:gaussianDemapper}
\end{equation}
with $z_{m,n,i}$ being the equalization error. Let the size of constellation $\Qc$ be $M = 2^K$ for some positive integer $K$. The Gaussian demapper acts on each $\hat{x}_{m,n,i}$ separately with the assumption that the error $z_{m,n,i}$ is Gaussian noise. With this assumption, the \gls{LLR} for the $j^{th}$ bit of User $i$ on RE $(m,n)$ is
\begin{align}
    LLR(m,n,i,j) = \textrm{log} \LB \frac{\sum_{c \in C(j,0)}P(c|\hat{x}_{m,n,i})}{ \sum_{c \in C(j,1)}P(c|\hat{x}_{m,n,i})}\RB, 
\end{align}
$\forall i =1,\cdots, N_u$, $\forall j=1,\cdots,K$, where $C(j,b)$ is the set of all constellation symbols with bit $b$ in the $j^{th}$ position. Due to the Gaussian assumption on $z_{m,n,i}$, $P(c|\hat{x}_{m,n,i})$ is computed as 
\begin{equation}
   P(c|\hat{x}_{m,n,i}) \propto \exp{-\frac{|c-\hat{x}_{m,n,i}|^2}{r_{mn}^{(i)}}}
\end{equation}
where $r_{mn}^{(i)}$ is the $(i,i)^{th}$ element of $\Rm_{z,m,n}$. The LLRs are then sent to the \gls{LDPC} decoder.

\begin{figure*}[htbp]
    \centering
    \includegraphics[width=1.\textwidth]{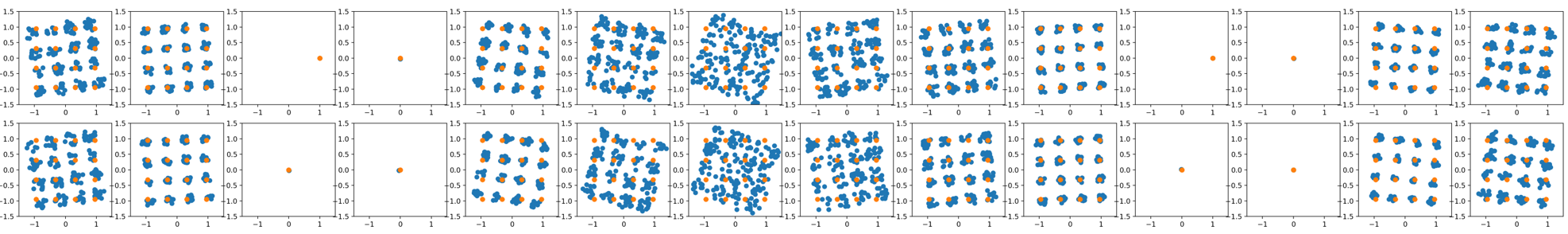}
    \caption{Plots of samples of the equalized signal for one of the users on 2 subcarriers and 14 \gls{OFDM} symbols. The orange points are the actual transmitted symbols from $16$-QAM, and the blue points are the samples of the equalized signals.}
    \label{fig:motivation}
\end{figure*}

\subsection{Motivation for a Neural Demapper}
\label{sec:motivation}

 With the nearest-pilot channel estimator, the assumption that the error $z_{m,n,u} $ is Gaussian distributed is no longer a reasonable one. Moreover, the equalization errors are correlated across time and frequency, as illustrated in Fig. \ref{fig:motivation}. In the figure, we have plotted samples of the equalized signals $\hat{x}_{m,n,i}$ for two subcarriers and $14$ \gls{OFDM} symbols, and for one of the $4$ co-scheduled users with the pilot pattern as shown in Fig. \ref{fig:operation}. From this figure and from \eqref{eq:lmmse_equalization}--\eqref{eq:eqNoiseVar}, the errors of each user exhibit the following properties. 
 \begin{enumerate}
    \item There is a relationship between the errors of adjacent REs for the same user, as shown in Fig. \ref{fig:motivation}.
    \item The errors of each user are dependent on the amount of correlation between the user's channel and the channels of other users (inter-user interference).
 \end{enumerate}
Let $\hat{\Xm}_i \in \CC^{N_f \times N_t}$ denote the grid of equalized signals for User $i$ so that the $(m,n)^{th}$ entry of $\hat{\Xm}_i$ is $\hat{x}_{m,n,i}$ (given by \eqref{eq:gaussianDemapper}). Similarly, let $\Rm_{X,i} \in \RR^{N_f \times N_t}$ denote the matrix whose $(m,n)^{th}$ entry is $r_{mn}^{(i)}$ which is the $(i,i)^{th}$ element of $\Rm_{z,m,n}$. Let $\hat{\Xm}  \triangleq [\hat{\Xm}_1^T, \cdots, \hat{\Xm}_{N_u}^T]^T \in \CC^{N_u \times N_f \times N_t}$, and $\Rm_{X}  \triangleq [\Rm_{X,1}^T, \cdots, \Rm_{X,N_u}^T]^T \in \RR^{N_u \times N_f \times N_t}$. So, we are interested in finding a suitable function 
 \begin{eqnarray} \nonumber
     f : \CC^{N_u \times N_f \times N_t} \times \RR^{N_u \times N_f \times N_t}  & \longrightarrow & \RR^{N_u \times N_f \times N_t \times K} \\  \label{eq:funtion_approx}
               f\LB \hat{\Xm}, \Rm_{X} \RB & \longmapsto & \Lm
 \end{eqnarray}
 where $\Lm \triangleq [\Lm_1^T, \cdots, \Lm_{N_u}^T]^T$, with $\Lm_i$ being the 3-dimensional array of LLRs for each bit of User $i$ in each RE of the grid. Since $f$ is not straightforward to obtain, we make use of techniques from machine learning to approximate it, and this is detailed in the following section.

\section{Convolutional Attention-based Neural Demapper}
\label{sec:NN}

In order to approximate the function given in \eqref{eq:funtion_approx}, we need a neural network that is capable of capturing the relationship between the errors of a user within the \gls{OFDM} grid. A natural choice for this would be a \gls{CNN}. In recent years, \gls{CNN}-based architectures have shown great results for the physical layer (see \cite{honkala2021deeprx, DBLP:journals/corr/abs-2009-05261}), and these use ResNet blocks \cite{DBLP:journals/corr/HeZRS15}. As corroborated by simulation results in Section \ref{sec:evaluations}, ResNet-based demappers are performance-limited for our task since the correlation between users is not captured. Indeed, each user's inputs are independently processed, and it is unlikely that the \gls{NN} learns to deal with this correlation. We need to capture the effects of the inter-user interference (caused by the correlation between the channels of the users). This is the main motivation for using self-attention that was originally proposed in \cite{vaswani2017attention} in the context of natural language processing.

 We draw inspiration from RE-MIMO \cite{RE_MIMO}, which uses self-attention for \gls{MU-MIMO} detection. In RE-MIMO, the encoder-decoder-based \gls{MHA} mechanism that was proposed in \cite{vaswani2017attention} is adapted to the physical layer. An inherent issue with using attention is that because it has its roots in natural language processing, the inputs need to be vectorized. If we wish to adapt it to a grid like in the case of OFDM-based signal processing or image processing, we end up with a very large amount of trainable parameters. There have been a few noteworthy efforts that address this issue for the task of image-classification. Some examples are \cite{ViT} and \cite{DBLP:journals/corr/swin} which implement a fully attention-based image classifier with a performance similar to that of CNN-based models. More recently, jointly using convolution and attention (convolutional-attention) \cite{DBLP:journals/corr/CvT, DBLP:journals/corr/LeFF} has shown great results, and we use these concepts for our purpose.

 \begin{table*}[htbp]
    \resizebox{\textwidth}{!}{
    \begin{tabular}{|l|l|l|l|l|l|}
    \hline
    \textbf{Layer}                   & \textbf{Type}  & \textbf{$\#$ Filters}                              & \textbf{Filter size} & \textbf{Dilation} & \textbf{Output dimension} \\ \hline\hline
    Input 1:  $\hat{\Xm} $ & Equalized  symbols         &  N/A                    & N/A                    &  N/A                         & $(B, N_u, N_f,N_t, 2)$                        \\ \hline
    Input 2: $\Rm_{X} $                         & Error Covariance &  N/A            & N/A                       &  N/A                          & $(B, N_u, N_f,N_t, 1)$                       \\ \hline
    Concat: $\Zm $                       & Concatenation of Inputs $1$ and $2$    &  N/A             &  N/A                      &   N/A                        & $(B, N_u, N_f,N_t, 3)$                     \\ \hline
    Reshape $\Zm$                       & Reshape     &  N/A            &  N/A                     &   N/A                         & $(BN_u,N_f,N_t, 3)$                     \\ \hline
    $Conv_{in}$                         & Conv2D     &  64                         & $(3,3)$                  & $(1,1)$                     & $(BN_u,N_f,N_t, 64)$                      \\ \hline
    Three CvT blocks                   & CvT         &  64      & $(3,3) $                 & $(1,1)$                     & $(BN_u,N_f,N_t,64)$                      \\ \hline
    $Conv_{out}$                        & Conv2D  & $K$                             & $(1,1)$                  & $(1,1)$                     & $(BN_u,N_f,N_t, K)$                      \\ \hline
    Output LLR                       & Transpose and Reshape        &  N/A                                  &  N/A                   &  N/A                        & $(B,N_f,N_t, N_u, K)$                               \\ \hline
    \end{tabular}
    }
    \caption{Convolutional attention-based demapper architecture}
\label{tab:cvt}
\end{table*}

\begin{figure}[htbp]
    \centering
    \includegraphics[width=0.45\textwidth]{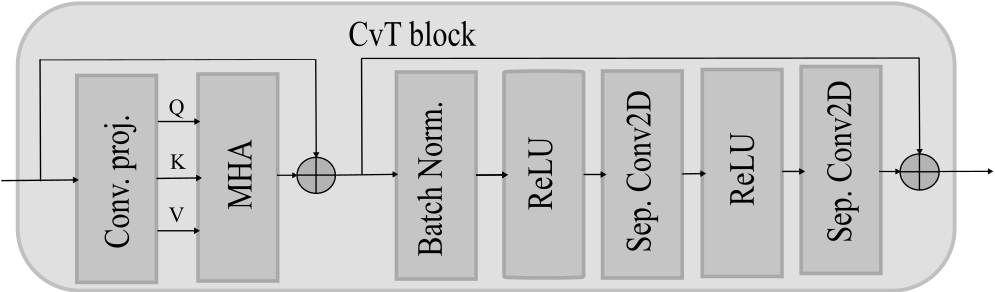}
    \caption{Convolutional Transformer block}
    \label{fig:CvT}
\end{figure}

\subsection{Model Architecture}

Attention, as the name suggests, aims to learn the degree of emphasis to place on the other users' equalized signals. It is done by generating \gls{K}, \gls{Q} and \gls{V} from the inputs, and more details on how they work is available in \cite{vaswani2017attention}. The building block of our \gls{NN} is the \gls{CvT} block shown in Fig. \ref{fig:CvT}. It is characterized by two parameters: $d_k$, the inner dimension of the model, and $N_h$, the number of multi-heads. "Multi-Head Attention" is used to obtain the values of \gls{K}, \gls{Q}, and \gls{V} from the inputs. The usage of attention in \gls{MU-MIMO} detection is shown to yield a very good performance in \cite{RE_MIMO}. It is shown that self-attention is able to focus entirely on users belonging to the same cluster, and hence the ones which have correlated channels. This is what motivated us to use attention in our proposed \gls{NN}.

\begin{algorithm}
    \caption{The sequence of operations for the \gls{NN} of Table \ref{tab:cvt}}
     \label{alg:NN}    
     \KwIn{$\hat{\Xm} \in \RR^{B\times N_u  \times N_f \times N_t \times 2 }$, $\Rm_{X} \in \RR^{B\times N_u  
     \times N_f \times N_t \times 1 }$, $d_{m}=64$}       
     \KwOut{$\Lm \in  \RR^{B\times N_u  \times N_f \times N_t \times K }$}
     $\Zm \gets \mathrm{Concatenate}\LB \hat{\Xm}, \Rm_{X} \RB \in  \RR^{B \times N_u  \times N_f \times N_t \times 3 } $\;
     $\Zm \gets \mathrm{Reshape}\LB \Zm \RB \in \RR^{BN_u\times N_f \times N_t \times 3} $\;
     $\Zm \gets \mathrm{Conv2D}\LB \Zm \RB \in \RR^{BN_u\times N_f \times N_t \times d_{m}} $\;
     $\Zm \gets \mathrm{CvT}\LB \Zm \RB \in \RR^{BN_u\times N_f \times N_t \times d_{m}} $\;
     $\Zm \gets \mathrm{CvT}\LB \Zm \RB \in \RR^{BN_u\times N_f \times N_t \times d_{m}} $\;
     $\Zm \gets \mathrm{CvT}\LB \Zm \RB \in \RR^{BN_u\times N_f \times N_t \times d_{m}} $\;
     $\Zm \gets \mathrm{Conv2D}\LB \Zm \RB \in \RR^{BN_u\times N_f \times N_t \times K} $\;
     $\Lm \gets \mathrm{Reshape}\LB \Zm \RB \in \RR^{B \times N_u\times N_f \times N_t \times K} $\;
\end{algorithm}

\begin{algorithm}
    \caption{The sequence of operations for the \gls{CvT} block of Fig. \ref{fig:CvT}}
     \label{alg:CvT}    
     \KwIn{$\Zm \in \RR^{BN_u\times N_f \times N_t \times d_{m}}$}       
     \KwOut{$\Zm \in  \RR^{BN_u  \times N_f \times N_t \times d_{m} }$}
     $\Qm \gets \mathrm{SeparableConv2D}\LB \Zm \RB \in  \RR^{B N_u  \times N_f \times N_t \times d_{m} } $\; 
     $\Qm \gets \mathrm{BatchNorm}\LB \Qm \RB \in  \RR^{B N_u  \times N_f \times N_t \times d_{m} } $\;   
     $\Qm \gets \mathrm{Rearrange}\LB \Qm \RB \in  \RR^{B N_f N_t \times N_u \times d_{m} } $\; 
     $\Km \gets \mathrm{SeparableConv2D}\LB \Zm \RB \in  \RR^{B N_u  \times N_f \times N_t \times d_{m} } $\; 
     $\Km \gets \mathrm{BatchNorm}\LB \Km \RB \in  \RR^{B N_u  \times N_f \times N_t \times d_{m} } $\;   
     $\Km \gets \mathrm{Rearrange}\LB \Km \RB \in  \RR^{B N_f N_t \times N_u \times d_{m} } $\; 
     $\Vm \gets \mathrm{SeparableConv2D}\LB \Zm \RB \in  \RR^{B N_u  \times N_f \times N_t \times d_{m} } $\; 
     $\Vm \gets \mathrm{BatchNorm}\LB \Vm \RB \in  \RR^{B N_u  \times N_f \times N_t \times d_{m}} $\;   
     $\Vm \gets \mathrm{Rearrange}\LB \Vm \RB \in  \RR^{B N_f N_t \times N_u \times d_{m} } $\; 
     $\Am \gets \mathrm{MHA}\LB \Qm, \Km, \Vm \RB \in \RR^{B N_f N_t \times N_u \times d_{m} } $\;
     $\Am \gets \mathrm{Rearrange}\LB \Am \RB \in \RR^{BN_u \times N_f \times  N_t \times d_{m} } $\;
     $\Zm \gets \Zm + \Am \in \RR^{BN_u \times N_f \times  N_t \times d_{m} } $\;
     $\Wm \gets \mathrm{BatchNorm}\LB \Zm \RB\in \RR^{BN_u \times N_f \times  N_t \times d_{m} } $\;
     $\Wm \gets \mathrm{SeparableConv2D}\LB \mathrm{ReLU}\LB \Wm \RB \RB\in \RR^{BN_u \times N_f \times  N_t \times d_{m} } $\;
     $\Wm \gets \mathrm{SeparableConv2D}\LB \mathrm{ReLU}\LB \Wm \RB \RB\in \RR^{BN_u \times N_f \times  N_t \times d_{m} } $\;
     $\Zm \gets \Zm + \Wm \in \RR^{BN_u \times N_f \times  N_t \times d_{m} } $\;
\end{algorithm}

\begin{algorithm}
    \caption{The sequence of operations for the \gls{MHA} block of Fig. \ref{fig:CvT}}
     \label{alg:MHA}    
     \KwIn{$\Qm, \Km, \Vm \in \RR^{B N_f N_t \times N_u \times d_{m}}$, $d_k=8, N_h = d_m/N_h = 8$}       
     \KwOut{$\Am \in  \RR^{B N_f N_t \times N_u \times 64}$}
     $\Qm \gets \mathrm{Dense}\LB \Qm \RB \in  \RR^{B N_f N_t \times N_u \times d_{m} } $\; 
     $\Qm \gets \mathrm{Reshape}\LB \Qm \RB \in  \RR^{B N_f N_t \times N_u \times N_h \times d_k } $\; 
     $\Qm \gets \mathrm{Transpose}\LB \Qm \RB \in  \RR^{B N_f N_t \times N_h \times N_u \times d_k } $\;   
     $\Km \gets \mathrm{Dense}\LB \Km \RB \in  \RR^{B N_f N_t \times N_u \times d_{m} } $\; 
     $\Km \gets \mathrm{Reshape}\LB \Km \RB \in  \RR^{B N_f N_t \times N_u \times N_h \times d_k } $\; 
     $\Km \gets \mathrm{Transpose}\LB \Km \RB \in  \RR^{B N_f N_t \times N_h \times N_u \times d_k } $\;
     $\Vm \gets \mathrm{Dense}\LB \Vm \RB \in  \RR^{B N_f N_t \times N_u \times d_{m} } $\; 
     $\Vm \gets \mathrm{Reshape}\LB \Vm \RB \in  \RR^{B N_f N_t \times N_u \times N_h \times d_k } $\; 
     $\Vm \gets \mathrm{Transpose}\LB \Vm \RB \in  \RR^{B N_f N_t \times N_h \times N_u \times d_k } $\;   
     $\Am \gets \mathrm{Softmax}\LB \frac{\Qm\Km^T}{\sqrt{d_k}}\RB \Vm \in \RR^{B N_f N_t \times N_h \times N_u \times d_k } $\;
     $\Am \gets \mathrm{Rearrange}\LB \Am \RB \in \RR^{B N_f N_t \times N_u  \times N_h \times d_k } $\;
     $\Am \gets \mathrm{Reshape}\LB \Am \RB \in \RR^{B N_f N_t \times N_u  \times d_{m} } $\;
     $\Am \gets \mathrm{Dense}\LB \Am \RB \in \RR^{B N_f N_t \times N_u  \times d_{m}  } $\;
\end{algorithm}

The architecture of the proposed convolutional attention-based \gls{NN} is depicted in Table \ref{tab:cvt}. The sequence of operations of the proposed \gls{NN} is shown in Algorithm \ref{alg:NN}. The sequence of operations specific to the \gls{CvT} block is shown in Algorithm \ref{alg:CvT} while that of the \gls{MHA} block is shown in Algorithm \ref{alg:MHA}. In these algorithms, $B$ refers to the batch size, and "Rearrange" refers to multiple transpose and reshape operations rather than just a simple reshape.

\subsection{Training}

Let $\Gc$ denote the set of all RE index pairs so that $\vert \Gc \vert = N_f N_t$, and let $b_{m,n,i,j} \in \{0,1\}$ denote $j^{th}$ bit transmitted by the $i^{th}$ user on the $(m,n)^{th}$ RE. Then, following the notation used in the previous section, the training aims to maximize the rate 
\begin{align}
\label{eq:rate}
     R &=  \frac{1}{N_u\vert \Gc \vert}\sum_{i = 1}^{N_u}\sum_{(m,n)\in \Gc}\sum_{j=1}^K I(b_{m,n,i,j};\hat{\Xm},\Rm_{X}) \nonumber\\
    & - \frac{1}{N_u\vert \Gc \vert}\sum_{i = 1}^{N_u}\sum_{(m,n)\in \Gc}\sum_{j=1}^K \mathbb{E}_{\hat{\Xm}, \Rm_{X}}\Big[ \nonumber \\ 
    & D_{KL}\big[q(b_{m,n,i,j}|\hat{\Xm},\Rm_{X})||p(b_{m,n,i,j}|\hat{\Xm},\Rm_{X})\big]\Big] 
\end{align}
where $R$ is an achievable information rate \cite{bocherer2018achievable}, $I(X;Y)$ the mutual information between random variables $X$ and $Y$, $D_{KL}(q||p)$ the Kullback–Leibler divergence between distributions $q$ and $p$, $q(b_{m,n,i,j}|\hat{\Xm},\Rm_{X})$ the conditional posterior distribution of $b_{m,n,i,j}$ generated by the demapper, and $p(b_{m,n,i,j}|\hat{\Xm},\Rm_{X})$ corresponds to the true posterior distribution. It is straightforward to calculate $q(b_{m,n,i,j}|\hat{\Xm},\Rm_{X})$ from the \gls{LLR} generated by the demapper. The first term in \eqref{eq:rate} corresponds to the rate achieved by an optimal demapper ($q =p$) while the second term can be viewed as the rate-loss caused by an imperfect demapper.

Let $\Theta$ denote the set of trainable parameters of our learned demapper. We use \gls{SGD} to optimize it, and it has been shown in \cite{DBLP:journals/corr/abs-2009-05261} that maximizing $R$ in \eqref{eq:rate} is equivalent to minimizing the \gls{BCE} between $q$ and $p$. Therefore, we generate a batch (of size $B$) of \gls{iid} equiprobable bits $\{b_{m,n,i,j}^{(s)}\}_{s=1}^{B}$, $i=1,\cdots,N_u$, $j=1,\cdots,K$, $(m.n) \in \Gc$ in each training epoch. Let $q_{\Theta}(b_{m,n,i,j}^{(s)}) \triangleq q_{\Theta}\LB b_{m,n,i,j}^{(s)}|\hat{\Xm}^{(s)},\Rm_{X}^{(s)}\RB$, the conditional distribution of $b_{m,n,i,j}^{(s)}$ generated from the output \gls{LLR} of the demapper. Using the system model in Section \ref{sec:problem_positioning}, we train the \gls{NN} by using the following loss function on this batch.

\begin{align}
    L(\Theta) & \triangleq \frac{1}{B N_u\vert\Gc\vert}\sum_{s = 1}^B\sum_{i = 1}^{N_u}\sum_{(m,n)\in \Gc}\sum_{j=1}^K \mathrm{log}_2\LB q_{\Theta}(b_{m,n,i,j}^{(s)}) \RB
\end{align}

The channel is generated using QuaDRiGa \cite{Jaeckel2014} in order to obtain the equalized signals in the above equation.

\section{Simulation Results} 
\label{sec:evaluations}

Two set of channel realizations were generated using QuaDRiGa channel generation tool according to the 3GPP UMi LOS/NLOS model. We used $50,000$ realizations for training and $10,000$ realizations for evaluation. Since we evaluate the performances for an \gls{SNR} in the range \SI[input-quotient  = :, output-quotient = \text{--}]{10:31}{\dB}, the training was done in the range \SI[input-quotient  = :, output-quotient = \text{--}]{7:34}{\dB}. Also, during the training the randomly generated bits were not LDPC-encoded, the reason being that the randomness of the samples can be beneficial and can potentially speed up the training. Table \ref{tab:parameters} summarizes the main parameters of the simulation. It took $200,000$ iterations for the training loss to satisfactorily converge for the proposed \gls{NN}. We also used a ResNet-based demapper (the 3 \gls{CvT} blocks replaced by 5 ResNet blocks) for comparison in order to highlight the significance of the attention mechanism. This ResNet-based demapper converged within $30,000$ iterations.

\begin{table}[htbp]
    \begin{center}
        \begin{tabular}{|l|c|}
        \hline
            {\bf Parameter}      & {\bf Value}                         \\ \hline \hline
            $N_t$  & 14     \\ \hline
            $N_f$   & 72  \\ \hline
            $N_u$   & 4  \\ \hline
            $N_r$   & 16    \\ \hline
            Sub-carrier spacing $\Delta f$ (\si{kHz})    & 15  \\ \hline
            Cyclic prefix duration $\Delta CP$ (\textmu s)     & 6    \\ \hline              
            $(d_{m}, d_k, N_h)$   & $(64,8,8)$     \\ \hline
            Batch size $B$ & $2$ \\ \hline
            LDPC code-rate  $r$   & $1/2$ \\ \hline
            User Speed (\si{\kmph})  & $3.6$    \\ \hline
            Modulation   & QPSK \\ \hline
        \end{tabular}
    \end{center}
    \caption{Simulation Parameters}
    \label{tab:parameters}
\end{table}

Figure \ref{fig:ber} presents the \gls{BER} for the following receivers -- a genie-aided receiver with perfect \gls{CSI} and Gaussian demapping (optimal in this case), the nearest-pilot channel estimator with a Gaussian demapper (baseline), the ResNet-based demapper, and our proposed \gls{CvT}-based demapper. Compared to the baseline, the ResNet-based demapper doesn't provide any significant gain. On the other hand, the \gls{CvT}-based demapper provides gains of around \SI{10}{\dB} compared to the baseline and is much closer to the perfect \gls{CSI} scheme. 

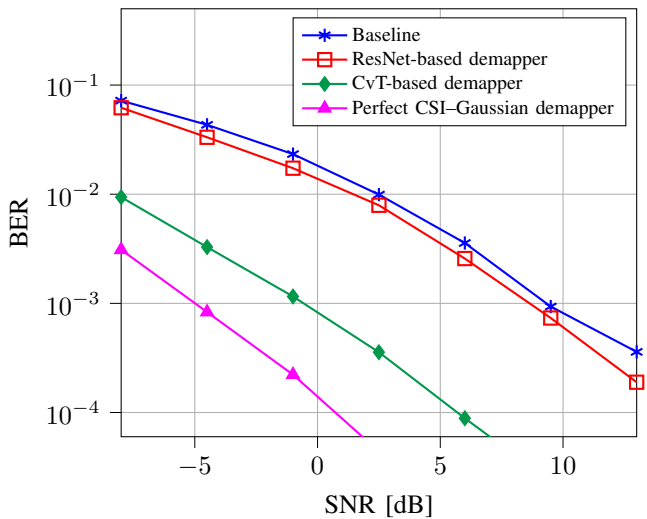
\begin{figure}[htbp]
\centering
\begin{tikzpicture}

\definecolor{color0}{rgb}{0, 0, 1}
\definecolor{color1}{rgb}{1,0,0}
\definecolor{color2}{rgb}{0, 0.6, 0.298039215686275}
\definecolor{color3}{rgb}{1,0,1}

\begin{axis}[
legend cell align={left},
legend style={nodes={scale=0.75, transform shape}}, 
log basis y={10},
tick align=outside,
tick pos=left,
x grid style={white!69.0196078431373!black},
xlabel={SNR [dB]},
xmajorgrids,
xmin=-8, xmax=13,
xtick style={color=black},
y grid style={white!69.0196078431373!black},
ylabel={BER},
ymajorgrids,
ymin=6e-05, ymax=0.5,
ymode=log,
ytick style={color=black},
ytick={1e-06,1e-05,0.0001,0.001,0.01,0.1,1,10},
yticklabels={
  \(\displaystyle {10^{-6}}\),
  \(\displaystyle {10^{-5}}\),
  \(\displaystyle {10^{-4}}\),
  \(\displaystyle {10^{-3}}\),
  \(\displaystyle {10^{-2}}\),
  \(\displaystyle {10^{-1}}\),
  \(\displaystyle {10^{0}}\),
  \(\displaystyle {10^{1}}\)
}
]
\addplot [thick, color0, mark=asterisk, mark size=2.5, mark options={solid}]
table {%
-8 0.0719719752669334
-4.5 0.0432301536202431
-1 0.0233072638511658
2.5 0.00991629995405674
6 0.00357251078821719
9.5 0.000934163574129343
13 0.000359456287696958
};
\addlegendentry{Baseline}
\addplot [thick, color1, mark=square, mark size=2.5, mark options={solid}]
table {%
-8 0.0619719752669334
-4.5 0.0332301536202431
-1 0.0173072638511658
2.5 0.00791629995405674
6 0.00257251078821719
9.5 0.000734163574129343
13 0.000189456287696958
};
\addlegendentry{ResNet-based demapper}
\addplot [thick, color2, mark=diamond*, mark size=2.5, mark options={solid}]
table {%
-8 0.00940040219575167
-4.5 0.00327803078107536
-1 0.00115610437933356
2.5 0.000356738368282095
6 8.85741537786089e-05
9.5 2.32910369959427e-05
13 6.44531473881216e-06
};
\addlegendentry{CvT-based demapper}
\addplot [thick, color3, mark=triangle*, mark size=2.5, mark options={solid}]
table {%
-8 0.0030989577062428
-4.5 0.000832519261166453
-1 0.000222167960600927
2.5 4.39453033322934e-05
6 0
9.5 0
13 0
};
\addlegendentry{Perfect CSI--Gaussian demapper}
\end{axis}

\end{tikzpicture}
\caption{\gls{BER} comparison}
\label{fig:ber}
\end{figure}

\section{Concluding Remarks}
\label{sec:conclusion}

In this paper, we presented a convolutional attention-based demapper for \gls{MU-MIMO} detection in the presence of a very simple channel estimator. This demapper was shown to be capable of significantly reversing the adverse effects of imperfect channel estimation. The gains provided by the proposed \gls{NN} over the baseline are very promising, and it could be interesting to study the usage of attention in other physical layer applications for \gls{MU-MIMO} systems. Optimizing the number of trainable parameters of this proposed \gls{NN} in order to make it more practical could be another direction of research.


\bibliographystyle{IEEEtran}
\bibliography{IEEEabrv, bibliography}

\end{NoHyper}
\end{document}